# The four elements in Robert Grosseteste's
# De Impressionibus Elementorum

**Amelia Carolina Sparavigna**
Department of Applied Science and Technology, Politecnico di Torino, Italy

*In De Impressionibus Elementorum, a treatise written by Grosseteste shortly after 1220, we can find a discussion of some phenomena involving the four classical elements (air, water, fire and earth), in the framework of an Aristotelian physics of the atmosphere. However this treatise strongly differentiates from similar works of the twelfth century for its referring to experiments. In fact, it contains some descriptions of phase transitions which are rather interesting, in particular when Grosseteste is discussing of bubbles.*

Robert Grosseteste (c. 1175 – 1253) was an English scientist and philosopher, bishop of Lincoln from 1235 AD until his death. As told in [1], he was commentator and translator of Aristotle and other Greek thinkers, and among the several scientific short works he wrote, we may find several references to the Aristotelian physics.

In the De Iride [2], Grosseteste is explicitly citing one of the Aristotelian works, that on Meteorology. This treatise contains Aristotle's theories about the earth sciences, including accounts of water evaporation, weather phenomena, and earthquakes. Let us remember that the Aristotelian physics was based on the four classic elements (Air, Water, Fire and Earth), to which he added the aether which is making the heavens. Aristotle is then describing in his Metereology a spherical lithosphere (Earth), a hydrosphere (Water) and the atmosphere (Air and Fire), surrounding them. He considered that the vapor which is formed during the day rises in the atmosphere to form the clouds, however not too high. "One reason for this is that it rises from hollow and watery places, so that the heat that is raising it, bearing as it were too heavy a burden cannot lift it to a great height but soon lets it fall again." [3] Not surprisingly, Aristotle is considered the father of climatology and geophysics [4].

After this short remark on the Aristotelian ideas on earth and atmosphere, we can read one of the scientific treatises written by Grosseteste, entitled De Impressionibus Elementorum. This title is usually translated as "On the Impressions of the Elements". "impressio" in Latin means "assault, impetus, vehemence," and then figuratively "perception"; then another translation of this title could be "On the Impetus of Elements." According to R.C. Dales, in this treatise, written shortly after 1220, "the main features of his scientific method are clearly in evidence", and it strongly differentiates from similar works of the twelfth century [5]; we find in several of the treatises written by Grosseteste a constant referring to experiments, which is used here too.

At the beginning of the XIII century, some of the medieval scholars were strongly influenced by the Aristotle's philosophy, which had begun to circulate in France in Arabic translation, introduced from Spain [6,7]. With this Aristotelian revival, we have a reassessment of using the fours classical elements in discussing physics, to feature the simplest principles of which anything is consisting. Most frequently, these classical elements refer to the phases of matter and then the Earth is solid, Water is liquid, Air is gas, and Fire is heat. Even in the poetry and religious songs of the XIII century we find the four elements to describe the Creation, such as in the Canticle of Sun, composed by Francis of Assisi [8].

As we will see reading De Impressionibus Elementorum, Grosseteste is discussing how an element can be changed, for instance the ice in water and the water in vapor, by using heat and fire. Quite interesting is the discussion on bubbles.

Here in the following I report the Latin text, as given in Ref.9. The text is subdivided in several sections. I am giving a translation and some notes, in the same manner as made for the De Iride and the De Generatione Sonorum [2,10-12]. A translation of this work was also given by R.C. Dales [5]; in some parts, my translation is enhancing physics.

INC: Ut testatur Jacobus in canonica sua "omne datum optimum et omne donum perfectum desursum est descendens a patre luminum, apud quem non est transmutatio, nec vicissitudinis obumbratio". EXPL: Differt autem pruina a nube, sicut differt pluvia a rore.

*INC: As told by James in his letter (James.1.17) "Every best thing and every perfect gift is coming down from the Father of Lights, with whom there is no mutation or shadow of change". END: hoar-frost is different from a cloud, such as rain differs from dew.*

When Grosseteste was bishop of Lincoln, he used to end the treatises writing "Explicit tractatus secundum Lincolniensem". Here, there is not this sentence: it means that the treatise was written before 1235, shortly after 1220 [5].

1 – Ut testatur Jacobus in canonica sua "omne datum optimum et omne donum perfectum desursum est descendens a patre luminum, apud quem non est transmutatio, nec vicissitudinis obumbratio". Hoc autem fieri in quibusdam rebus immediate, in quibusdam mediate necesse est. Quare philosophi, etsi perfecte res non intelligentes, cum naturas rerum non ignorare debent, radios corporum supercaelestium descendentes super res corporales mutationis earum maximam causam praebere non ignorantes dicunt, quod radii reflexi et condensati causa sunt caloris generati apud nos. Cuius signum est, quod in convallibus maior calor est, quam in montibus; unde in montibus diutius manet nix, quam in convallibus; unde etiam in quibusdam montibus altissimis manet nix perpetua.

*As told by James in his letter (James.1.17) "every best thing and every perfect gift is coming down from the Father of Lights, with whom there is no mutation or shadow of change". However, this is resulting, under some circumstances, immediately, and in other cases, needs mediation. Then philosophers, even if they are not perfectly able of understanding the facts, ought not to be ignorant of the nature of things, and they do not ignore that the rays of the heavenly bodies falling upon the physical things provide the greatest cause of their changes, because the rays when reflected and condensed are the cause of heat generated among us. A proof of this is the fact that the heat is greater in the valleys than on the mountains; and then snow remains longer on the mountains than in the valleys; and on some high mountains snow remains perpetually.*

2 – Et nota quod nihil differt, quod sol in se sit calidus. Si enim corpus solare in se esset calidum et calor eius excitaret calorem in rebus inferioribus, tunc quanto propinquiores essent ei res, tanto calidiores essent et in cacumine montium esset maior calor, quam in convallibus et in superiori et medio interstitio aeris esset maior, quam in infimo; cuius oppositum ex toto videmus, quia in cacuminibus montium manet nix, in convallibus autem non; et in interstitio superiori generantur grandines et in infimo pluviae. Signum ad idem: aves rapaces in aestate ascendunt, ut frigefiant, ut aquilae volant in sublime, ut temperent calorem suum generatum ex motu; multum enim volant. Et grues et aliae aves multae in convallibus descendunt contra forte gelu, contra vero calidum ascendunt ad montana. Et haec omnia eiusdem signa sunt, scilicet quod calor non provenit ex corpore solari, sed ex reflexione et condensatione radiorum.

*And let us note that it makes no difference, that the sun is hot in itself. For if the body of the sun were considered as hot in itself and its heat exciting heat in the things below it, then the closer a thing were to it the warmer it would be, and on the tops of the mountains there would be greater heat than in the valleys and in the upper and medium layers of air more heat than in the lower layer; but we see all the opposite facts, because the snow remains on the tops of the mountains, not in the valleys; and in the upper layer of air the hail is generated and in the lower the rain. A sign of the same: the birds of prey in the summer fly high to cool themselves, like the eagles flying very high, to mitigate the heat generated from the movement; they fly so much. The cranes and many other birds come down in the valley down to escape ice and frost, on the other hand, to escape hot climate they go up to the mountains. And all these are signs of the same, that is, that heat is not transferred directly from the solar body, but from the reflection and condensation of rays.*

We know that there are three methods by which heat is transferred: conduction, convection, and radiation. Conduction and convection are supported by solid and fluid media. But they cannot account for some of other phenomena, for instance, the heat we feel when sitting in front of a fire or under the sun. And then is seems that Grosseteste knew this aspect of the heat transfer, in particular by radiation, as distinguished by conduction.
In this section we find that Grosseteste subdivides the atmosphere in three layers according to Aristotle.

3 – Hoc habito manifestum est, quod radii in aquis descendunt ad fundum, cum aqua sit corpus transparens sicut aer, cornuglacies et vitrum. In fundis ergo aquarum est reflexio, quare maior est calor in fundo, quam in superficie. Unde pisces in hieme in fundo aquarum sunt, in aestate vero in superficie et congelatur aqua in superficie, in fundo vero non.

*Under these conditions it is clear that the rays go down deep in the water, being the water a transparent medium such as air, icicle and glass. Therefore, some deviation of rays exists in deep waters, then, the heat is greater at the bottom than at the surface. Hence, at the bottom of waters, the fishes live during the winter, but, in summer, they can live near the surface; during the winter, the water is frozen at the surface, however, not in its depths.*

I translated "cornuglacies" as "horn of ice" and then icicle. In fact, I imagined the Grosseteste saw the icicles as horns of ice. In [5] it is used "hornglass", however this word is not mentioned in Latin vocabularies (for instance Gastiglioni-Mariotti, Loescher Editor).
Grosseteste invokes a "reflexio" for the thermal behavior of water. In fact, this Latin word means "turning away or back". Therefore I softened the translation, and instead of using reflection, I preferred a deviation of rays. In the De Iride, Grosseteste is discussing more deeply the light and the phenomena of reflection and refraction [2].
As a generic comment to this section, let us tell that Grosseteste observed a temperature gradient in the water of lakes and linked the temperature of water to the life in the lakes. However, the behavior of the thermal gradient in a lake is rather different [13].

4 – Si autem quaerat quis, quare aqua multum frigefacta congelatur, cum frigiditas sit eius naturalis potentia, similiter et humiditas ut videtur et fluxibilitas, ad hoc respondendum est, quod omnis aqua naturaliter est frigida, sed non fluida, immo potius ex natura sua est congelata. Fluxibilitas autem eius est ex calore incluso, ut mollities in terra.

*If anyone asks the following, why the water congeals when it is very cold, with the coldness being its natural power, as it seems to be humidity and fluidity, we answer to this person that all the water is naturally cold, but not fluid, by its nature it is frozen indeed. The fluidity results from the heat absorbed, for softening the land.*

According to Aristotle in his On Generation and Corruption, the element Water was primarily cold and secondarily wet. To understand the distinction in primary and secondarily features, let us remember that Air is primarily wet and secondarily hot, Fire is primarily hot and secondarily dry and the Earth is primarily dry and secondarily cold. That is, we have four elements and four features (wet, dry, hot, cold), to describe the natural phenomena.
Here, Grosseteste introduced the phase transition from solid to liquid, the melting of ice, and the fact that the transition is due to the heat absorbed. The phase transition occurs due to a change in energy of the participating particles. If the water is in the solid phase and the kinetic energy of molecules is sufficiently increased, we change the solid to liquid. In the solid phase, the molecules prefer to assume the lowest energy assembly. After the transition in the liquid phase, the total energy is larger. Let us say that Grosseteste argued that "cold" means a lower energy state of a substance.

5 – Item radii reflexi a speculo concavo generant ignem ut stupa apposita inflammetur. Habito ergo, quod calor proveniat ex condensatione radiorum manifestum est, quod, dum condensantur in fundis aquarum, calefit aqua et tantum calefit, quod non remanet sub natura aquae: transit igitur ad naturam aeris. Sed cum natura aeris non sit esse sub aqua, ascendit super aquam; ascendit autem in ampulla ex ipsa aqua. Multae autem ampullae, cum ascendunt super aquam, simul se tenent per naturam humiditatis ampullarum et ex illis fit vapor vel fumus, ex quibus fiunt nubes. Sed cum in fundo aquarum sit generatio ampullarum, quaedam earum transeunt per meatus terrae, quaedam remanent in aquis, quaedam ascendunt super aquam.

*Again, the rays reflected from a concave mirror generate fire and tinder is ignited. Therefore, having established that the heat is clearly coming from a condensation of rays, we have that, being them condensed in the bulk of water, the water is heated, and even heated so much that it does not keep its nature of water: it passes, therefore, to the nature of the air. But, to the nature of the air it is not proper being under the water, it comes out, over the water; but it rises in a bubble as in an ampoule made of the same water. However, when several bubbles ascend on the water, due to the nature of their wet films, barely can they remain themselves, and from them, vapor or steam is formed, by which the clouds are made. But, when the generation of bubbles is in the depths of the waters, some of these bubbles pass through the earth, some remain in the waters, and some rise above the water.*

It is interesting to note that Grosseteste is not using the word "bulla" for bubble, but he prefers "ampulla", ampoule, which is a small glass vial. It means that he observed that these bubbles were objects which were spherically contained volumes of vapors, made from a liquid. Rising at the surface of water, the wet films forming the bubbles break and the vapor inside them creates a cloud of steam, as we can see in the following section, where Grosseteste describes the boiling water.

6 – De ascendentibus primo dicendum est. Si quis autem sensibiliter velit videre, ponat aquam claram in patella clara et videbit manifeste ampullas generatas et ascendentes per calorem ignis suppositi sub patella. Idem enim est modus generationis ampullarum hic et ibi.

*Let us first talk about what is rising. If anyone wants to see this directly, put some clear water in a clear vessel, and you will perceive clearly the bubbles generated and rising, raised by the heat of the fire being placed under the vessel. Then, the same mode of generation of bubbles we have here (in this section) and there (that is, in the section 5).*

Here there is a Grosseteste's referring to experiment.

7 – Notandum tamen, cum aere et cum ampulla esse terram et ignem. Sunt ergo ibi quattuor elementa in ista ampulla scilicet terrestreitas propter locum generationis, aer generatus, natura ignis in generatione caloris; de aqua manifestum est. Haec ergo est quasi prima generatio elementorum et prima admixtio elementorum. Ubi est ergo aqua abundans in ampullis generatis, cum ascendunt super aquam, vocatur vapor humidus; quando terra est abundans, vocatur fumus siccus; cum vero aer sit abundans erit vapor pinguis. Vapor ergo ascendens ascendit super quantitatem et grossitiem et subtilitatem caloris generati. Si enim calor fuerit magnus et grossus, generatur ampulla magna et grossa et ponderosa. Unde quandoque non ascendit, nisi ad superficiem aquae et ibi insensibiliter frangitur et calor evaporatur. Et quanto subtilior est calor, et subtilior est ampulla et tunc debilis est calor.

*We have to note yet that with air and bubble there are earth and fire. In the bubble, therefore, there are four elements, that is, the earth because of the place of generation, the air which is generated, the nature of fire during the generation of heat, and of course some water. Then, here we find a sort of first generation of the elements and the first mixture of them. When there is an abundance of water in the generated bubbles, that is, when they go up on the water, we call it "humid vapor"; when the earth is abundant, we call it "dry smoke"; when the air is predominant, we call it "dense vapor". Therefore, the rising vapor rises according to the quantity, coarseness and subtlety of the generation of heat. If the heat is great and coarse, the generated bubble is great and coarse and heavy. Then, sometimes, it is not rising or just to the surface of water and there breaks imperceptible and the heat evaporates. And when the heat is more subtle, the bubble is more subtle and then weaker the heat (that evaporates).*

It is suitable to remember that evaporation and boiling are different, and surely Grosseteste noted the difference, because he tried to distinguish the great and coarse bubbles from the very subtle ones. The boiling of a liquid happens at the boiling point of it, that is, at the temperature at which its vapor pressure equals the environmental pressure. At the boiling point, the vapor pressure overcomes the atmospheric pressure and it is allowed the bubbles to grow in the bulk of the liquid and rise. However, liquids may change to a vapor at temperatures below their boiling points through the process of evaporation. Evaporation is a surface phenomenon in which molecules escape outside the liquid as vapor, without bubbles; boiling is a bulk process in which molecules escape, resulting in the formation of vapor bubbles within the liquid.
Besides boiling, bubbles are present in the so-called effervescence process, which is the process creating the sparkling wines. It is the result of the interplay between $CO_2$-dissolved gas molecules, tiny air pockets trapped within microscopic particles during the pouring process, and liquid properties. Ref. [14] summarizes the physicochemical processes behind the nucleation, rise, and burst of bubbles found in glasses poured with sparkling wines.

However, even tap water produces some bubbles too. The water has air dissolved in it. The amount of air that can be dissolved increases with pressure but decreases with temperature. Water in the tubes is usually colder than room temperature, and then the solubility of air in it is higher: as the water is poured, it warms up and the solubility of air is reduced. The air comes out even creating some bubbles. In fact, the air solute molecules can cluster together to form nuclei. When these nuclei are trapped by some defects on the glass surfaces, they start growing forming bubbles in the solution. Experimenting with tap water we see a slow formation and growth of bubbles, but if carbonated water is used, due to the excess of $CO_2$-dissolved gas, the bubbles form and grow rapidly.

Supposing that Grosseteste observed the evaporation of water and the contemporary formation of the bubbles from the air dissolved in it, or even the bubbles in wine, he could have argued that evaporation was accompanied by the formation of very subtle "bubbles".

8 – Unde non differt ampulla a superficie terrae et volitant hinc inde in convallibus. Hoc autem fit in vespertinis et in matutinis temporibus, quando debilior est calor et sic generatur nebula. Et cum illae ampullae parvae destruuntur a calore, cadunt ad superficiem terrae et fit ros. Si autem maior sit calor, facit istas ampullas praedictas sive nubem ascendere ad primum interstitium aeris. Est enim primum, secundum et tertium. Tertium vero non est altitudinis maioris, quam quinguaginta milliariorum, ut dicit philosophus.

*Then, bubbles do not separate from the earth surface and float here and there in the valleys. However, this occurs in the evening and in the morning, when the heat is weak, and so mist is formed. And when these small bubbles dissipate their heat, fall on the earth surface and create dew. However, if the heat is greater, it makes these aforementioned bubbles to rise at the first layer of the atmosphere. There are a first, a second and a third layer in the atmosphere. The third, however, is not at a height greater than fifty miles, as the Philosopher says.*

The Philosopher is Aristotle.
Grosseteste describes the mist or fog forming from the surface of water bodies, and connects it with the evaporation. The fog is due to the vapor that condenses into tiny liquid droplets in the air. Observing these small droplets in the fog, Grosseteste imagined their origin from vapor through its "bubbles".

9 – Cum ergo nubes sit in primo interstitio, destruitur quandoque a calore et unaquaeque ampulla, cum sit in nube, trahit se ad sui profundum. Et ideo separantur ampullae ab invicem et minutae guttatim cadunt. Immo fiunt guttae etsi nubes sit continua et quia non ex toto destruitur, a calore cadit fluida pluvia et non congelata. Et nota, quod generatio pluviae et roris differunt secundum magnum et parvum et secundum diversa loca generationis.

*Then, being the clouds in the first layer, sometimes each bubble loses its heat, and, since it is in the cloud, moves itself in the depth of the cloud. And then, the bubbles separate from each other and fall as small droplets. Drops occur indeed, although the cloud is continuous, and because it had not entirely deprived of heat, the rain falls, fluid and not frozen. And let us note that the generation of rain and dew differs according to a great or little size, and according to the different places of generation.*

When vapor rises in the atmosphere, we have a cloud, a visible mass of liquid droplets or frozen crystals.

10 – Cum vero ascendit nubes ad secundum interstitium, fit major abstractio caloris et destruuntur ampullae penitus a calore successive tantum, quare molle est quod relinquitur sicut lana et fit nix. Si autem subito deferatur nubes sursum ad secundum interstitium, subito destruitur a calore et fit lapis rotundus et est generatio grandinis, sicut ampulla fuit rotunda. Hoc autem fit maxime, cum calor fuerit magnus. Differt autem pruina a nube, sicut differt pluvia a rore.

*However, when the cloud rises to the second layer, there is a further loss of heat, and, the bubbles are then left utterly deprived of their heat only successively, and for this reason we have that they remain soft as wool, and become snow. However, if the cloud is suddenly rising to the second layer, suddenly is the heat lost and round stones, as the bubbles round, appear and hail is generated. This occurs especially when it is hot. However, hoar-frost is different from a cloud, such as rain differs from dew.*

As previously told, in this treatise, Grosseteste is following the Aristotle's model of the atmosphere, which is subdivided in three layers [15], in the lower and in the middles there are the clouds. Besides the meteorological phenomena, it is interesting the Grosseteste effort in describing the phases of matter, solid, liquid and vapor, and the transition between phases because of the involved heat. Moreover, it is remarkable that he considered the solid state of water as its fundamental state, being necessary some heat to gain its fluidity.